\documentclass[%
pra%
,reprint%
,twocolumn%
,secnumarabic%
,floatfix%
, showpacs%
,amssymb, amsmath, nobibnotes, aps,pra]{revtex4-1}

\usepackage[latin1]{inputenc}
\usepackage{epsfig}
\usepackage{color}
\usepackage{bm}
\usepackage{hyperref}

\newcommand{\bra}[1]{\left<#1\right|}
\newcommand{\ket}[1]{\left|#1\right>}

\begin{document}

\title{Coherent versus incoherent excitation dynamics in dissipative many-body Rydberg systems}
\author{David W. Sch\"{o}nleber}
\affiliation{Max-Planck-Institut f\"{u}r Kernphysik, Saupfercheckweg 1, 69117 Heidelberg, Germany}
\author{Martin G\"arttner}
\affiliation{Max-Planck-Institut f\"{u}r Kernphysik, Saupfercheckweg 1, 69117 Heidelberg, Germany}
\author{J\"org Evers}
\affiliation{Max-Planck-Institut f\"{u}r Kernphysik, Saupfercheckweg 1, 69117 Heidelberg, Germany}

\date{\today}

\begin{abstract}
We study the impact of dephasing on the excitation dynamics of a cloud of ultracold two-level Rydberg atoms for both resonant and off-resonant laser excitation, using the wave function Monte Carlo (MCWF) technique. We find that while for resonant laser driving, dephasing mainly leads to an increase of the Rydberg population and a decrease of the Mandel $Q$ parameter, at off-resonant driving strong dephasing toggles  between direct excitation of pairs of atoms and subsequent excitation of single atoms, respectively. These two excitation mechanisms can be directly quantified via the pair correlation function, which shows strong suppression of the two-photon resonance peak for strong dephasing.
Consequently, qualitatively different dynamics arise in the excitation statistics for weak and strong dephasing in off-resonant excitation. Our findings show that time-resolved excitation number measurements can serve as a powerful tool to identify the dominating process in the system's excitation dynamics.
\end{abstract}

\maketitle

\section{Introduction}

Ultracold gases of strongly interacting Rydberg atoms offer a rich playground for studying intriguing many-body phenomena reminiscent of spin
Hamiltonians long studied in solid state physics. The creation of self-assembled strongly correlated excitation structures has been studied for
resonant \cite{schauss2012,gaerttner2012,hoening2013,petrosyan2013b,ates2012b} as well as off-resonant
\cite{schempp2013,malossi2013,gaerttner2013a,robicheaux2005,weimer2010a,lemeshko2012,lee2011,mayle2011} laser driving. Moreover, the adiabatic
preparation of the crystalline ground states of these systems has been proposed, assuming fully coherent time evolution
\cite{pohl2010,bijnen2011,schachenmayer2010}. However, recent theoretical \cite{honer2011, petrosyan2013, petrosyan2013b, hoening2013, ates2012,
glaetzle2012, gorshkov2012, lee2011, lee2012, lee2012a, gaerttner2013} and experimental \cite{schempp2013,malossi2013} efforts have shown that
decoherence effects play a crucial role in the many-body dynamics of such systems.

The goal of this paper is to provide a detailed study of the influence of dephasing on the dynamics of resonantly as well as off-resonantly driven
ensembles of two-level Rydberg atoms. We provide an intuitive picture for understanding the relevant microscopic mechanisms in terms of the many-body
state space. With this picture at hand we identify signatures of coherent and incoherent dynamics in various observables using extensive numerical
simulations. This study closely relates to recent experimental findings \cite{schempp2013}.

For theoretical modeling of the quantum many-body dynamics of an ensemble of strongly interacting two-level atoms, the main obstacle to overcome is the problem of the exponential growth of the Hilbert space with the atom number. In the regime of coherent laser driving, an exact treatment via the Schr\"odinger equation is therefore limited to small atom numbers \cite{weimer2008, younge2009, robicheaux2005, gaerttner2012}.
In the strongly dissipative regime, in turn, the many-body master equation can be approximated by a rate equation (RE) \cite{ates2007,heeg2012} whose basis states grow only linearly with the atom number. Hence, an efficient simulation of large systems consisting of several thousand atoms is rendered possible.
In the weakly dissipative regime, neither rate equation nor Schr\"odinger equation are applicable, and a straightforward solution of the master equation for the density matrix is in general not feasible for systems consisting of several tens of atoms, due to the unfortunate scaling of the density matrix elements with the atom number. Consequently, the wave function Monte Carlo (MCWF) technique \cite{carmichael1991,dalibard1992,dum1992,molmer1993} has for lattice geometries recently been employed as an alternative but equivalent means to solve the many-body master equation \cite{lee2012,lee2012a,ates2012,petrosyan2013}. 

Here, we focus on two main questions, namely how dissipation (dephasing) alters the dynamics of a disordered, finite two-level Rydberg system, and which features can be used to uniquely distinguish between coherent and incoherent excitation dynamics. To answer these questions, we employ wave function Monte Carlo simulations.

We find that dephasing leads to a stronger saturation of the medium with Rydberg excitations for resonant driving and accordingly to lower Mandel $Q$ parameter values. In contrast, two qualitatively different kinds of excitation dynamics arise in off-resonant excitation, depending on the strength of the dephasing. In (quasi-)coherent off-resonant excitation, two-photon resonance effects dominate the dynamics and lead to a transient bimodality in the excitation number histogram between zero and two Rydberg excitations. In incoherent off-resonant excitation, sequential single-photon processes dominate, allowing in large systems temporarily for excitation number histograms that are characterized by a appreciable ground state $\ket{gg\cdots g}$ population and at the same time appreciable population of large excitation numbers. 
For the simulation of large systems we utilize a rate equation model which is benchmarked using MCWF technique.
Time-resolved study turns out to be of particular importance since key features in the observables only develop over time and the steady state of the system is not necessarily reached on typical experimental timescales.

\section{\label{sec:mcwf_model}Wave function Monte Carlo model}
Being based on an approach that solves the Schr\"odinger equation, the computational efficiency of our model relies on the state space truncation technique \cite{weimer2008, younge2009, robicheaux2005, gaerttner2012}, which ignores the dynamics of the states that are only negligibly populated due to their large energy generated by inter-atomic interactions. As state space truncation is most profitable for two-level systems consisting of a non-interacting ground state and a strongly interacting Rydberg state, we constrain the following discussion to two-level systems only. In three-level Rydberg systems, state space truncation can only be applied to the the strongly interacting Rydberg state, so the presence of a non-interacting intermediate state still leads to an exponential scaling of the state space with the atom number.

The two-level Rydberg systems we consider consist of a ground state $\ket{g}$ and an excited Rydberg state $\ket{r}$, e.g. the $\ket{50S_{1/2}}$ state of ${}^{87}$Rb, which interacts repulsively, with the interaction strength quantified by the van der Waals coefficient $C_6$. For an ensemble of $N$ atoms, labeled by $\alpha$, the Hamiltonian in rotating wave approximation reads ($\hbar=1$)
\begin{equation}
\mathcal{H}=\sum_{\alpha=1}^N \left[\mathcal{H}_\mathrm{L}^{(\alpha)}+\mathcal{H}_\Delta^{(\alpha)}\right] + \sum_{\alpha<\beta}\frac{C_6}{|\bm{R}_\alpha-\bm{R}_\beta|^6} \ket{r_\alpha r_\beta}\bra{r_\alpha r_\beta},
 \label{eq:Hamiltonian}
\end{equation}
where $\mathcal{H}_\mathrm{L}^{(\alpha)}=\Omega/2 \ket{g_\alpha}\bra{r_\alpha} + h.c.$ describes the coupling of the atoms to the laser field and $\mathcal{H}_\Delta^{(\alpha)}=-\Delta\ket{r_\alpha}\bra{r_\alpha}$ accounts for the detuning from the atomic transition frequency.

Starting from the von Neumann equation, incoherent processes such as dephasing or spontaneous emission can be included via Lindblad terms \cite{breuer2002}, leading to the master equation for the density matrix $\rho$,
\begin{equation}
\label{eq:masterequation}
 \dot{\rho}=-i[\mathcal{H},\rho]+\mathcal{L}[\rho].
\end{equation}

The incoherent processes we have in mind include a typically small effective decay of the Rydberg level $\ket{r_\alpha}$ quantified by the decay constant $\gamma$ as well as a decay of the coherences associated with the operator $\ket{g_\alpha}\bra{r_\alpha}$, quantified by the dephasing constant $\Gamma$. The considered dephasing might be due to the decay of the ignored intermediate state in a two-step excitation scheme \cite{petrosyan2013}, finite laser linewidth \cite{petrosyan2013}, collisional damping \cite{puri2001}, a speckle light field with short-range spatio-temporal correlations \cite{honer2011} or motional dephasing \cite{honer2011,gaerttner2013}.

In general, the MCWF technique \cite{carmichael1991,dalibard1992,dum1992,molmer1993} allows for the numerical solution of the many-body master equation \eqref{eq:masterequation} without employing the density matrix formalism.
Instead, a Schr\"odinger equation is solved and incoherent processes are included via non-Hermitian time evolution and additional random quantum jumps. Averaging over many such random trajectories allows one to obtain a mixed state that can be shown to solve the corresponding master equation. That way, the problem of solving a system of $n^2$ coupled ordinary differential equations can be reduced to a system of $n$ equations, where $n$ is the dimension of the underlying state space.

The wave function Monte Carlo algorithm proceeds as follows \cite{molmer1993}: In each step of the time evolution, the system can either undergo a coherent non-Hermitian time evolution or perform a quantum jump associated with incoherent processes such as dephasing or decay.
The non-Hermitian Hamiltonian including the anti-commutator part of the Lindblad term in Eq.~\eqref{eq:masterequation} is obtained via the replacement
\begin{equation}
	\mathcal{H} \rightarrow \mathcal{H}-\frac{i}{2}(\Gamma+\gamma)\sum_{\alpha=1}^{N}{\ket{r_\alpha}\bra{r_\alpha}}.
\end{equation}
The jump operators associated with dephasing and decay, respectively, are given accordingly by
\begin{equation}
	\mathcal{C}_\Gamma^{(\alpha)} = \sqrt{\Gamma}\ket{r_\alpha}\bra{r_\alpha}, \quad \mathcal{C}_\gamma^{(\alpha)} = \sqrt{\gamma}\ket{g_\alpha}\bra{r_\alpha}.
\end{equation}
Whether or not a quantum jump occurs in a single time step is determined by a Monte Carlo method, which randomly selects the jumping atom as well as the jump operator according to the jump probability given by  
\begin{equation}
	\delta p_{\alpha,\digamma} = \bra{\psi(t)} \mathcal{C}_\digamma^{(\alpha)\,\dagger}\mathcal{C}_\digamma^{(\alpha)} \ket{\psi(t)}\delta t,
\end{equation}
where $\psi(t)$ denotes the many-body wave function, $\digamma\in\{\Gamma,\gamma\}$ and $\delta t$ is chosen such that $\delta p = \sum_{\alpha,\digamma}{\delta p_{\alpha,\digamma}}\ll 1$. In addition, the first order approximation of the time evolution imposes another constraint on $\delta t$, requiring the modulus of the eigenvalues of the non-Hermitian Hamiltonian multiplied with the numerical time step to be small compared to $1$ \cite{molmer1993}.

As the norm of a wave function undergoing either non-Hermitian time evolution or a quantum jump is not conserved, the wave function is normalized after each time step, yielding a single (stochastic) wave function trajectory for each time evolution.
By averaging over the quantum trajectories, the master equation result is obtained in the limit of large sample size. Further details on the numerical modeling can be found in Refs.~\cite{gaerttner2012,gaerttner2013a,molmer1993}.

In Rydberg physics, the wave function Monte Carlo technique is most beneficial for calculations performed in the weakly dissipative regime, since in the strongly dissipative regime the computationally efficient rate equation model \cite{ates2007,heeg2012} proves reliable (cf. \cite{petrosyan2013a} and Sec.~\ref{subsec:RE_benchmark}).

\section{\label{sec:resonant}Dissipative excitation dynamics on resonance ($\Delta = 0$)}
Resonantly driven systems which interact via van der Waals interaction are characterized by the interaction-strength dependent blockade effect that manifests itself in the pair correlation function \cite{gaerttner2012,petrosyan2013a,petrosyan2013b}. Due to the avoided volume, the excitation number fluctuation is smaller as compared to Poissonian statistics obtained in independent excitation, resulting in a sub-Poissonian Mandel $Q$ parameter \cite{mandel1979}, $Q = \langle\left(\mathcal{N}-\langle\mathcal{N}\rangle\right)^2\rangle/\langle\mathcal{N}\rangle -1 < 0$. 
Here, we study how these effects are influenced by dissipation. We find that decoherence leads to an enhancement of the dense packing of Rydberg excitations, i.e., to lower values of the $Q$-parameter, and to a weakening of the blockade effect, i.e., to a smoother behavior of the pair correlation function. Moreover, differences are found in the temporal evolution of the system. In the presence of dissipation a slow increase in Rydberg population is observed which is not present in the coherent case. 

\subsection{\label{subsec:res_mechanisms}Basic excitation mechanisms}
Before discussing the results of our numerical calculations in detail, let us first outline the basic mechanism that we expect to be relevant for the formation of Rydberg excitations in the coherent and the incoherent case, respectively.

Considering a resonantly driven, fully blockaded ensemble of atoms (superatom), the population of the Rydberg level is significantly enhanced in the presence of dephasing due to an effect which we call \emph{superatom symmetry breaking} (SASB) \cite{honer2011,petrosyan2013a}. While for coherent driving, only the symmetric state (Dicke state) of the superatom couples to the ground state $\ket{gg\cdots g}$, this symmetry is broken in the presence of dephasing such that also the formerly dark states can be populated. SASB leads to a steady state excitation fraction of $N/(N+1)$ if the decay of the Rydberg level is neglected \cite{honer2011}. Note that while in the absence of spontaneous emission the steady state of the system does not depend on the value of the dephasing constant \cite{honer2011}, this does not hold true any more for nonzero decay.

Generalizing this reasoning to the case of an extended sample, permitting more than one excitation, it is clear that the loss of coherence will lead to an enhanced saturation of the excitation density, since the rate with which highly excited states are populated depends on their interaction energy. The broadening of the laser permits to access states with higher interaction energies (and hence smaller spacings between the excitations), which happens on a slower timescale than the initial dynamics. Thus one expects a slow increase of Rydberg population in the long time limit observed for incoherent resonant driving. We could confirm this mechanism by considering a two-atom system in which the equilibration timescale strongly depends on the interaction shift for weak decay and nonzero dephasing.

\subsection{\label{subsec:res_results}Results}

\begin{figure}[t]
  \centering
 \includegraphics[width=\columnwidth]{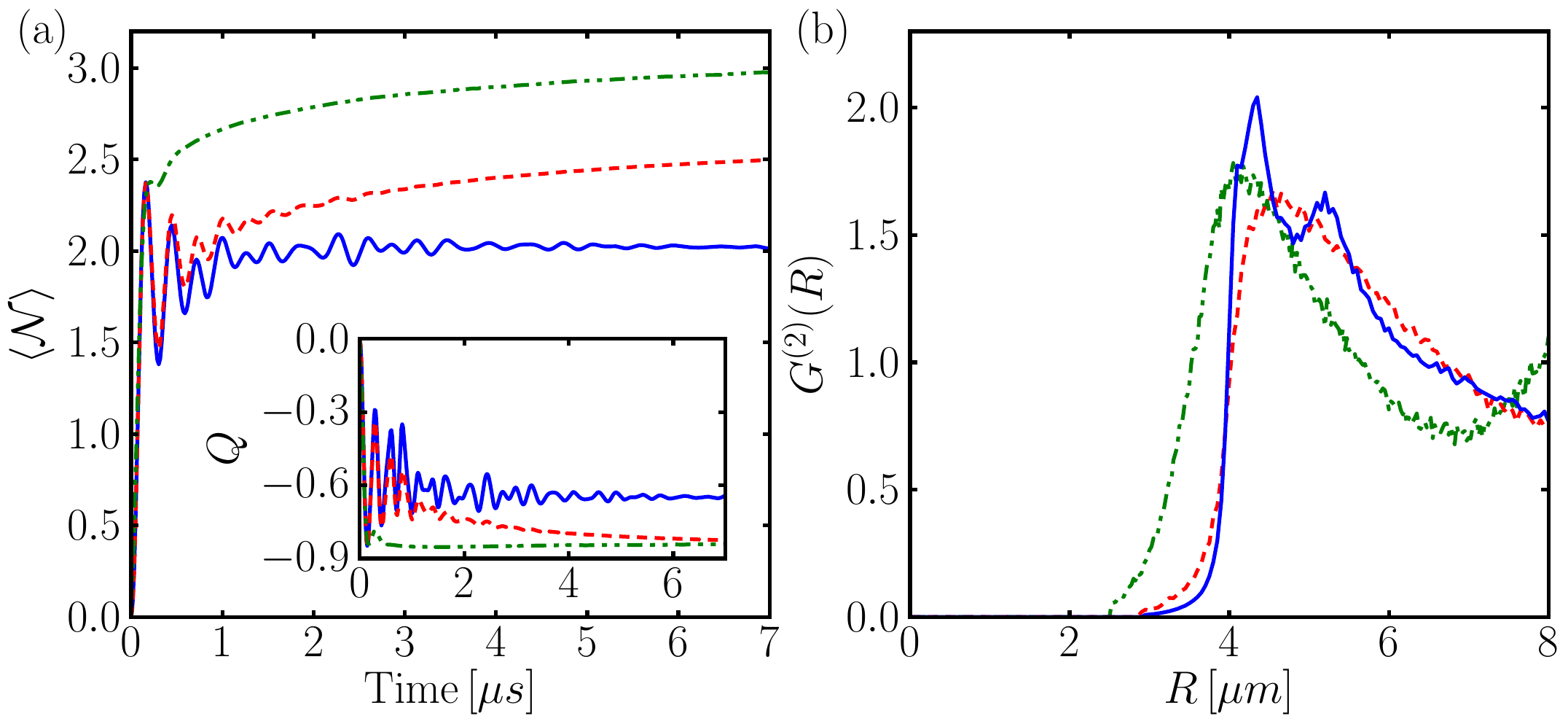}
 \caption{(Color online) $\mathrm{(a)}$ Mean Rydberg population $\langle\mathcal{N}\rangle$ as a function of time and $\mathrm{(b)}$ pair correlation function for different dephasings, $\Gamma=0$ (blue, solid), $\Gamma/2\pi=0.1$\,MHz (red, dashed) and $\Gamma/2\pi=2$\,MHz (green, dash-dotted). The inset in (a) shows the Mandel $Q$ parameter as a function of time. The pair correlation function is evaluated at $t=10\,\mu s$. Further parameters: $N=40$, $C_6/2\pi=16\,\mathrm{GHz}\,\mu m^6$, $\Omega/2\pi=0.8$\,MHz, $\gamma=0.025$\,MHz, $\Delta = 0$, one-dimensional trap of length $L_\mathrm{1D}=15\,\mu m$.}
 \label{fig:delta0_popuG2}
\end{figure}
To probe the conjectures of the previous section and to illustrate the mentioned mechanisms we carry out numerical simulations and analyze the dynamics of various observables. We first consider in Fig.~\ref{fig:delta0_popuG2}(a) the Rydberg population $\langle\mathcal{N}\rangle$ in a one-dimensional disordered geometry as a function of time. In the presence of dephasing, the population increases compared to the (quasi-)coherent case ($\Gamma=0$), with a slowdown of the population increase at large times.
This can be understood by means of the two mechanisms introduced above, namely the SASB and the dephasing-induced broadening of the laser linewidth. 
A similar slowdown was also observed in the limit of purely incoherent dynamics~\cite{lesanovsky2013}.
While the former directly effects an increase of the Rydberg population for nonzero dephasing, the latter allows for a slow population (even for large dephasing values) of Rydberg excited states in which the excitations feature a distance smaller than the blockade radius for zero dephasing.
The enhanced saturation of the excitation density expected for SASB is consistent with the dynamics of the $Q$ parameter shown in the inset of Fig.~\ref{fig:delta0_popuG2}(a), which displays reduced $Q$ values for nonzero dephasing.

Second, to further confirm the interpretations given above, we show the pair correlation function in Fig.~\ref{fig:delta0_popuG2}(b), defined as in \cite{heeg2012},
\begin{equation}\label{eq:G2_definition}
G^{(2)}(R) = \dfrac{\sum_{\alpha<\beta}{\langle\mathcal{N}^{(\alpha)}\mathcal{N}^{(\beta)}\rangle\delta_{R}^{(\alpha,\beta)}}}{\left(\frac{1}{N}\langle\mathcal{N}\rangle\right)^2\sum_{\alpha<\beta}{\delta_{R}^{(\alpha,\beta)}}}\,,
\end{equation}
where $\delta_{R}^{(\alpha,\beta)}$ is $1$ only if the two atoms $\alpha,\beta$ are separated by the distance $R$. The pair correlation function $G^{(2)}(R)$ quantifies the probability of finding two Rydberg excitations separated by the distance $R$. Since finite size effects dominate at the trap boundaries \cite{gaerttner2012}, we only show the first peak of $G^{(2)}(R)$. The double-peak structure of $G^{(2)}(R)$ for zero dephasing (solid blue curve) is also due to finite-size effects and should not be confused with the double-peak structure arising in off-resonant excitation because of resonance effects (cf. Sec.~\ref{subsec:smalltrap}), which are absent for $\Delta=0$. The finite-size origin of the double-peak structure for (quasi-)coherent excitation has been confirmed via additional simulations by increasing the trap size while holding the density constant, which showed that the peak heights strongly depend on the trap size and that the peaks merge for a large trap geometry. 
As can be seen from Fig.~\ref{fig:delta0_popuG2}(b), dephasing smears out the structures in $G^{(2)}(R)$ and weakens the blockade, which is indicated by the shift of the onset of $G^{(2)}(R)$ towards smaller distances for increased dephasing.
This weakening of the dipole blockade again confirms our interpretation of the slow increase of the Rydberg population at large times being due to the population of strongly-interacting states with more closely spaced excitations.

As a last observation we note that, as expected, the oscillations in both the population and accordingly the Mandel $Q$ parameter are damped out by the dephasing, with the damping rate being largest for largest dephasing.

\begin{figure}[t]
  \centering
 \includegraphics[width=\columnwidth]{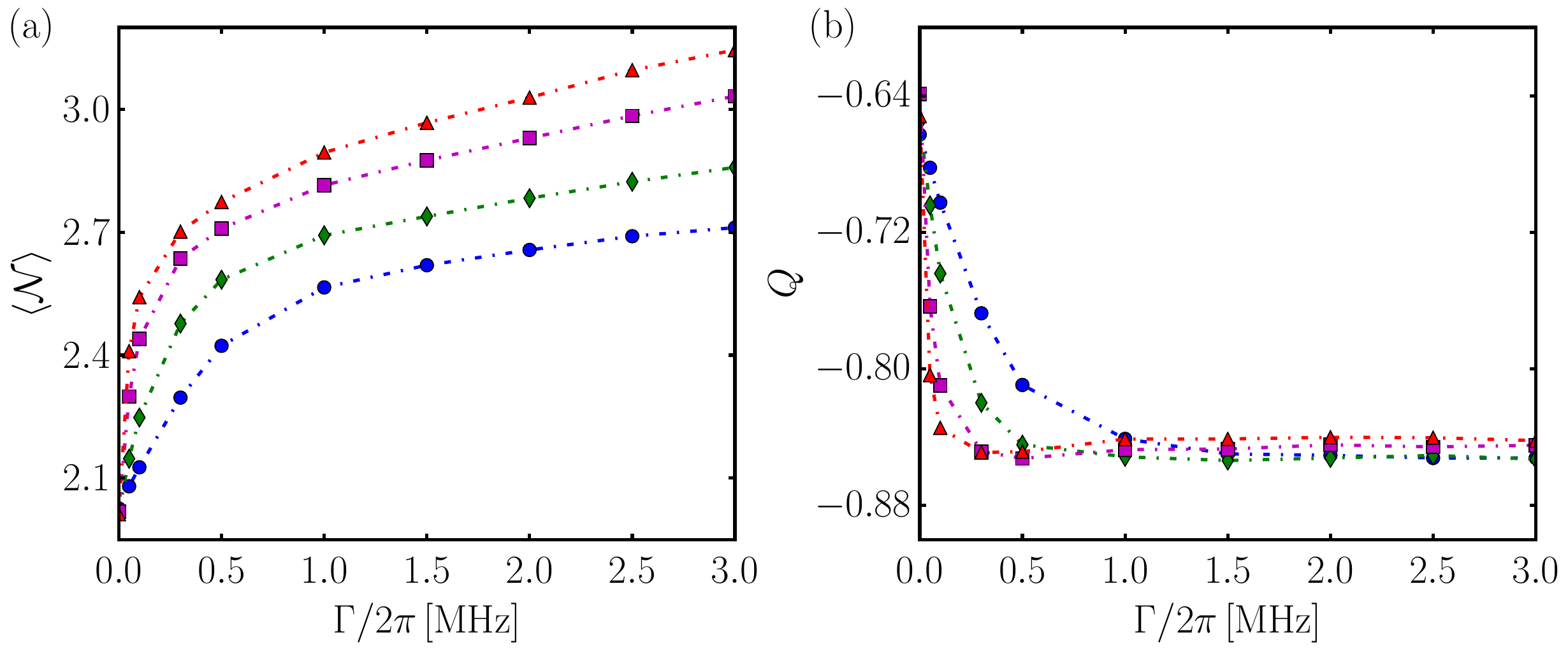}
 \caption{(Color online) $\mathrm{(a)}$ Mean Rydberg population $\langle\mathcal{N}\rangle$ and $\mathrm{(b)}$ Mandel $Q$ parameter as a function of the dephasing. The different symbols indicate different evaluation times, $t=\{1,2,5,10\}\,\mu s$, corresponding to the data points marked by circles, diamonds, squares, and triangles, respectively. Further parameters as in Fig.~\ref{fig:delta0_popuG2}.}
 \label{fig:delta0_popuQvsGamma}
\end{figure}

Next, we study the Rydberg population of the system for different evaluation times more systematically as a function of the dephasing in Fig.~\ref{fig:delta0_popuQvsGamma}(a). This shows that the Rydberg excitation indeed grows monotonically with the dephasing constant $\Gamma$, with the strongest relative population increase occurring when moving from a quasi-coherent system ($\Gamma=0$) to a weakly dissipative system (here, $\Gamma/2\pi = 0.05\,\mathrm{MHz}$). 
This behavior can likewise be understood from the aforementioned arguments, in particular the SASB mechanism, which explains the increase of $\langle\mathcal{N}\rangle$ for small dephasing values. 
Moreover, one observes the slow long-term population dynamics for large dephasings explained above.

\begin{figure}[t]
  \centering
 \includegraphics[width=\columnwidth]{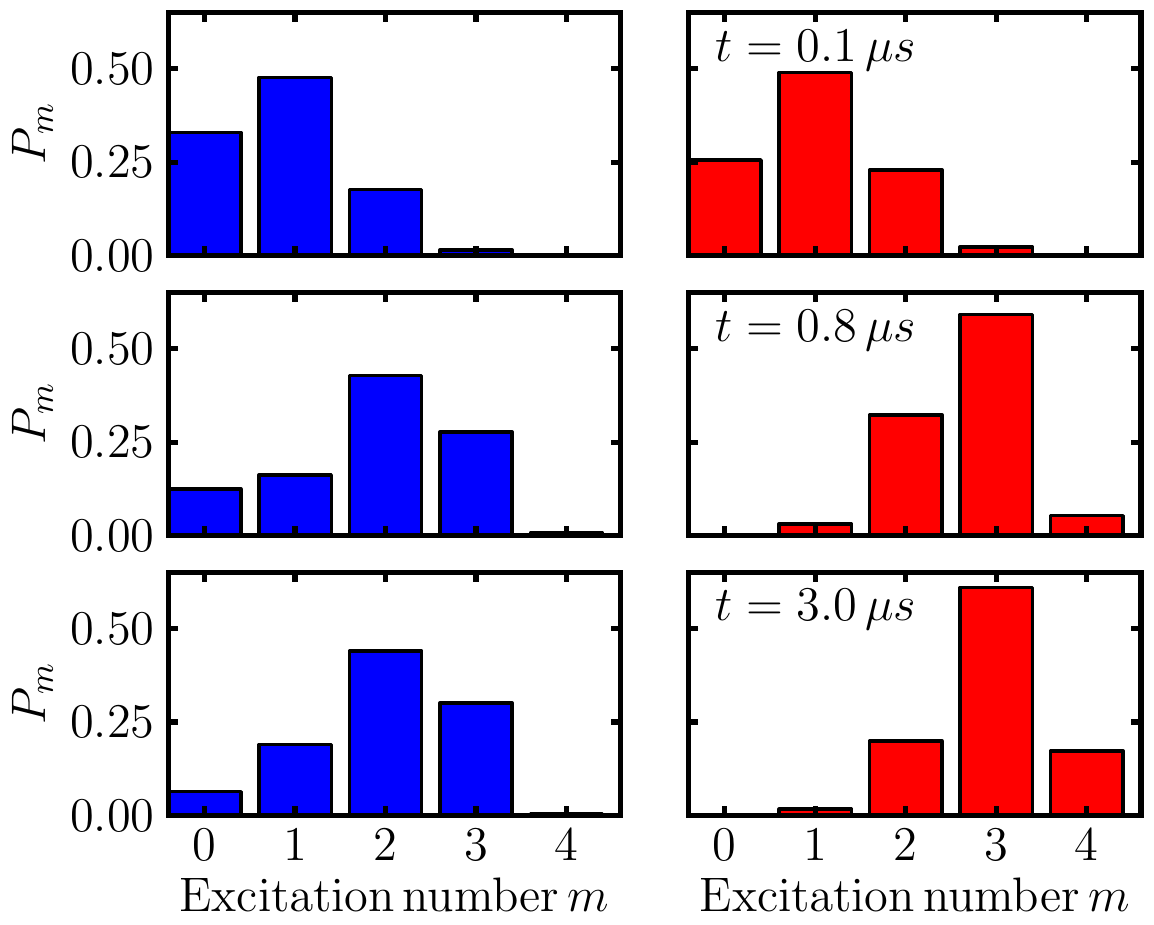}
 \caption{(Color online) Excitation probability $P_m$ for two different dephasings, $\Gamma=0$ (blue, left panel) and $\Gamma/2\pi=3$\,MHz (red, right panel), at three different times. Further parameters as in Fig.~\ref{fig:delta0_popuG2}.}
 \label{fig:delta0_hist_tdep}
\end{figure}

In addition, we consider the impact of dephasing on the Mandel $Q$ parameter, shown in Fig.~\ref{fig:delta0_popuQvsGamma}(b). Clearly, dephasing reduces the value of the $Q$ parameter compared to the quasi-coherent case, which, in a lattice geometry, has also been observed in \cite{petrosyan2013b}. This is a manifestation of dephasing-enhanced dense packing, i.e., the increase of the mean Rydberg population brings about a reduced fluctuation of the excitation numbers due to a strong saturation of the volume with Rydberg excited atoms. Again, $Q$ decreases most at the onset of the dephasing, but saturates subsequently. The increase of $Q$ at larger times for large dephasing is consistent with the interpretation that strongly-interacting states are slowly populated for large dephasing since the slow population of higher excited states leads to a (weak) broadening of the excitation number histograms at large evolution times, which is related to a larger (less sub-Poissonian) $Q$ value. The timescale on which the dense packing is established, i.e., the timescale of superatom dephasing, is proportional to $\Gamma$. This is why for small dephasing the dense packing is only achieved at long evolution times in Fig.~\ref{fig:delta0_popuQvsGamma}(b).

As an illustration of the dynamics of the excitation histograms, we show the probabilities $P_m$ for finding $m$ excitations at different times in Fig.~\ref{fig:delta0_hist_tdep}, comparing the quasi-coherent ($\Gamma=0$) with the strongly incoherent ($\Gamma/2\pi=3$\,MHz) case. In both simulations, $P_5$ is negligible, so the histograms are only shown up to $m=4$. Fig.~\ref{fig:delta0_hist_tdep} shows that for resonant laser driving the excitation process in the presence of dephasing is essentially characterized by the migration of ground state $\ket{gg\cdots g}$ population to higher excitation numbers via the intermediate excited state manifolds, leading to a (sub-Poissonian) distribution around the mean count. In the quasi-coherent case, the same picture applies, except for the more versatile short-term dynamics caused by population oscillations which lead to a temporary re-population of low excitation numbers due to Rabi oscillations.

We thus conclude that, for resonant excitation, the main effect of dephasing is to increase the mean Rydberg population by allowing the dark states of the superatoms to be populated and by weakening the dipole blockade, thereby reducing the fluctuations in the excitation numbers. A discrimination between coherent and incoherent features is difficult in the long time limit since the differences are only of quantitative nature. The strongest indicator of coherent time evolution is the collective Rabi oscillations at short times. Such oscillations for multiple atoms \cite{dudin2012} as well as for a single pair of atoms \cite{gaetan2009,beguin2013} have already been observed experimentally for fully blockaded ensembles.

\section{\label{sec:off_resonant}Excitation dynamics off resonance ($\Delta > 0$)}
Having studied the effect of dephasing in resonant excitation, we now turn to the discussion of dephasing-induced effects in off-resonant excitation. There, the detuning can compensate the energy shift induced by the Rydberg-Rydberg interaction [cf. Eq.~\eqref{eq:Hamiltonian}], giving rise to resonant excitation channels \cite{gaerttner2013a}. In this regime the system's properties are dominated by those resonant excitation mechanisms, manifesting itself in characteristic features in the pair correlation function and the excitation number statistics \cite{schempp2013}. Therefore it is crucial to understand the effects of dephasing on these mechanisms when studying off-resonant excitation in the dissipative regime.

\subsection{\label{subsec:offres_mechanisms}Basic excitation mechanisms}
For the off-resonant case, substantial differences between coherent and incoherent driving already arise in the case of a single atom. Fully coherent driving results in oscillations between the ground and the Rydberg state with the amplitude given by $\Omega^2/(\Omega^2+\Delta^2)$. The Rydberg population never exceeds this value. In the presence of dissipation, however, the system features a steady state,
\begin{equation}\label{eq:SS_Ryd_1a}
	\langle\mathcal{N}\rangle_\mathrm{SS} = \dfrac{1}{2+\gamma\frac{(4\Delta^2+(\gamma+\Gamma)^2)}{4(\gamma+\Gamma)\Omega^2}}\,.
\end{equation}
In the limit of a small spontaneous decay rate $\gamma$ (as typical of Rydberg systems) and nonzero dephasing, the steady state \eqref{eq:SS_Ryd_1a} approaches the value $1/2$ independent of the laser detuning, which means that the Rydberg fraction becomes significantly larger than the amplitude of the coherent oscillation. This already illustrates that for coherent driving (unitary time evolution) the achievable Rydberg excited fraction is determined by the transition matrix element of the initial state with any excited state, while for incoherent driving a picture of classical rates between the states is more adequate, where the achievable excitation depends on detailed balance relations between the states.

We employ this rate equation reasoning to quantify the timescales of (strongly) incoherent excitation dynamics. For a single atom, the excitation rate is given by  
\begin{equation}\label{eq:res_ex_rate}
	\gamma_{\uparrow} = \frac{\Omega^2(\gamma+\Gamma)}{4\Delta^2+(\gamma+\Gamma)^2}\,.
\end{equation}
Here, $\gamma_{\uparrow}$ is obtained from the single-atom master equation \eqref{eq:masterequation} by eliminating the atomic coherences via the condition $\dot{\rho}_{ij}=0$ for $i\neq j$ ($i,j \in \{g,r\}$) and solving for $\rho_{gr},\rho_{rg}$. Inserting $\rho_{gr},\rho_{rg}$ into the remaining equations and making further use of the trace condition $\mathrm{Tr}[\rho]=1$, the excitation rate $\gamma_{\uparrow}$ can be directly read off the equation for $\dot{\rho}_{rr}$ (cf. \cite{ates2007} for details). The resonant excitation rate $\gamma_{\uparrow,\mathrm{res}}$ is obtained by setting $\Delta=0$ in Eq.~\eqref{eq:res_ex_rate}, yielding in the limit $\Gamma\gg \gamma$ the rate $\gamma_{\uparrow,\mathrm{res}} = \Omega^2/\Gamma$.

\begin{figure}[t]
  \centering
 \includegraphics[width=\columnwidth]{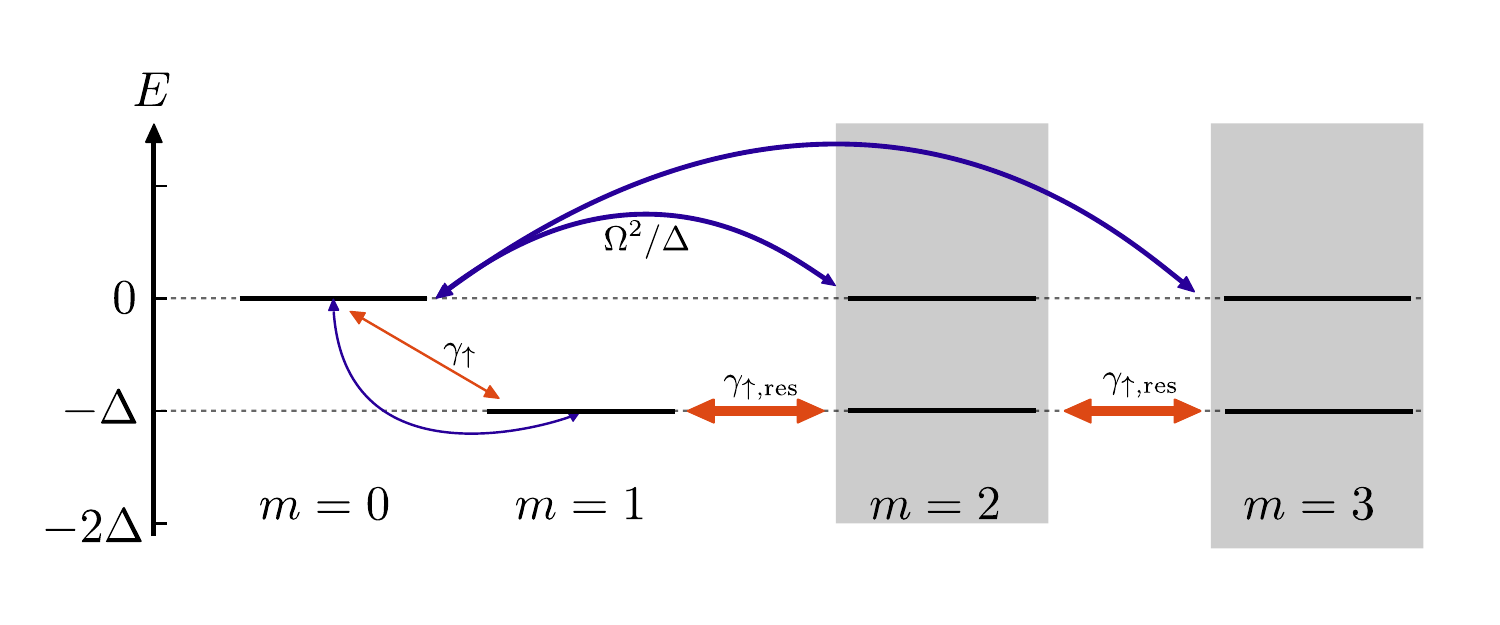}
 \caption{(Color online) Sketch of the dominant off-resonant excitation mechanisms in an illustrative few-excitations sample. The blue bent lines indicate coherent excitation, the red straight lines incoherent excitation. The thickness of the arrows quantifies the strength of the coupling. The grey boxes indicate the possible energies of the respective states in the rotating frame of Eq.~\eqref{eq:Hamiltonian} up to a certain energy cutoff.}
 \label{fig:excitation_mechanism}
\end{figure}

Resonant excitation can not only occur for zero detuning, but also when the Rydberg-Rydberg interaction compensates for the detuning. 
Consequently, the first excitation is rather unlikely for a large detuning $\Delta$ as $\gamma_{\uparrow}$ is small, yet subsequent excitations can be resonant if there are atoms in the appropriate single-photon resonance distance \cite{schempp2013}. Hence, we expect the excitation mechanism to be governed by two timescales: the slow timescale of the off-resonant initial (seeding) excitation and the fast timescale of resonant single-photon excitations, leading to subsequent growth of excitations (aggregates) after initial seed \cite{schempp2013}. This mechanism is illustrated in Fig.~\ref{fig:excitation_mechanism}, where straight double arrows represent incoherent excitation rates. Direct excitation of the multiply excited states is not possible (or strongly suppressed) since it requires inter-atomic coherences, which are strongly damped by the dephasing \cite{gaerttner2013a} (see also Sec.~\ref{subsec:RE_benchmark}). Thus, in a stepwise process, once a single excitation is present, further excitations can be created resonantly at $E=-\Delta$. 

In the coherent case, the classical picture of rates connecting the various many-body states does not apply. The time evolution is fully determined by the transition matrix element of the ground state $\ket{gg\cdots g}$ (initial state) with the various excited states, with dynamics arising from the superposition of the diverse couplings. As illustrated in Fig.~\ref{fig:excitation_mechanism}, the transition matrix element of the ground state with any far-detuned states is small. Thus the dynamics is restricted to a small energy band around $E=0$. For the couplings (bent double arrows) to the resonant ($E=0$) doubly excited states, a characteristic timescale can be identified, namely the effective two-photon Rabi frequency, which is given by $\Omega^2/\Delta$ \cite{gaerttner2013a}. It is obtained in a two-atom picture by adiabatically eliminating the singly excited states. Resonant couplings to higher excited states ($m>2$) are associated with even slower timescales \cite{schempp2013}. Identifying the direct excitation of doubly excited states as the dominant mechanism also means that a suppression of $P_{m=1}$ should be expected in the coherent case.

\begin{figure}[t]
  \centering
 \includegraphics[width=\columnwidth]{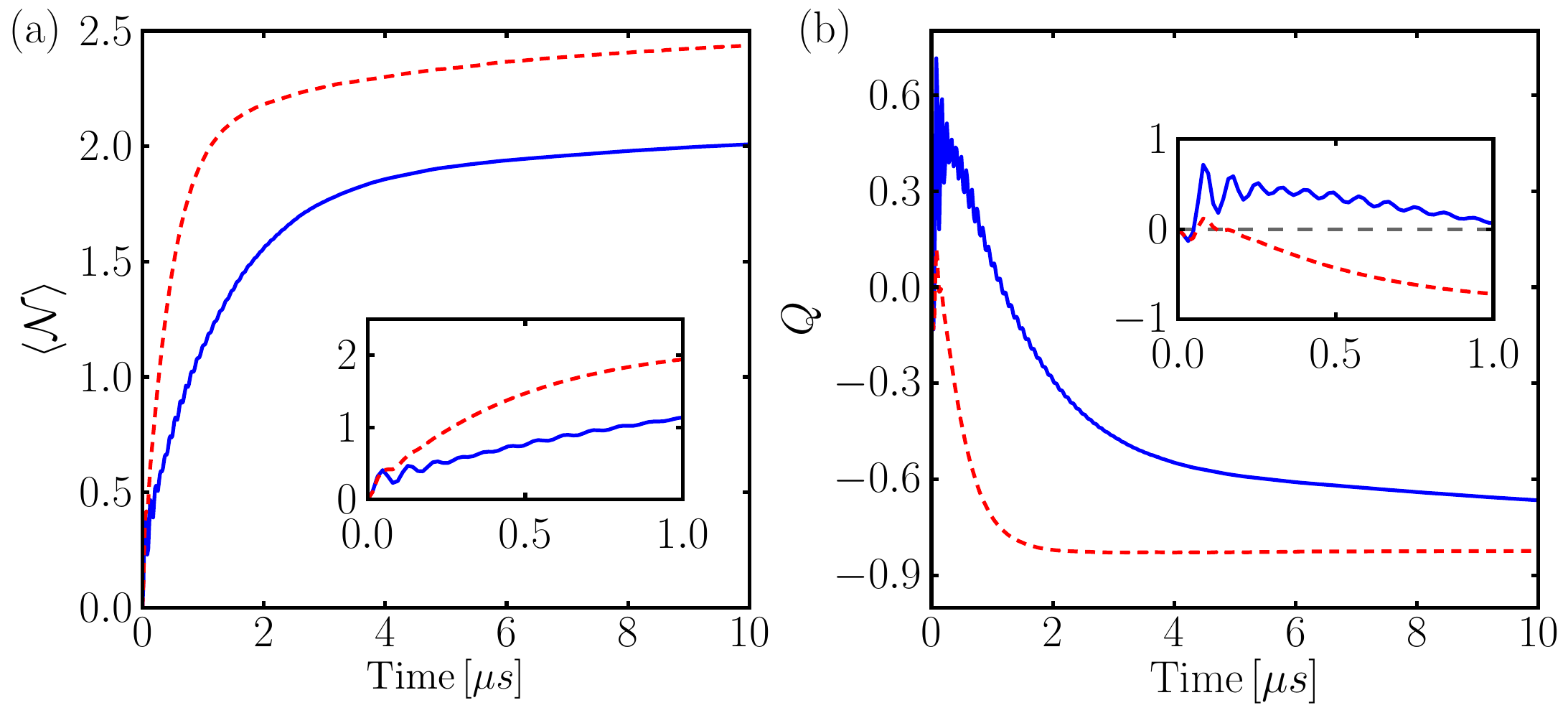}
 \caption{(Color online) $\mathrm{(a)}$ Mean Rydberg population $\langle\mathcal{N}\rangle$ and $\mathrm{(b)}$ Mandel $Q$ parameter as a function of time for different dephasings, $\Gamma/2\pi=0.05$\,MHz (blue, solid) and $\Gamma/2\pi=4$\,MHz (red, dashed). The insets show enlarged details of the respective short-term dynamics. Further parameters: $N=50$, $C_6/2\pi=16\,\mathrm{GHz}\,\mu m^6$, $\Omega/2\pi=1$\,MHz, $\gamma=0.025$\,MHz, $\Delta/2\pi = 10$\,MHz, cylindrical trap with radius $R = 1.65\,\mu m$ and length $L=6\,\mu m$.}
 \label{fig:WM_popuQ}
\end{figure}

In summary, for the coherent case only a narrow part of the state space with energies around $E=0$ is accessible due to the narrow two-photon resonance, which in addition leads to a slowdown of the dynamics at large detuning. In contrast, incoherent excitation leads to diffusion into a broad band of states with energies around $E=-\Delta$.

\subsection{\label{subsec:smalltrap}Small trap geometry}
For our numerical study of the effects of dissipation for off-resonant driving we switch to a three-dimensional geometry being closer to realistic experimental setups. We consider a cylinder of length $L=6\,\mu m$ and radius $R = 1.65\,\mu m$, in which $50$ atoms are randomly placed and evaluate the dynamics for two different dephasing constants, $\Gamma/2\pi=0.05$\,MHz and $\Gamma/2\pi=4$\,MHz, respectively. Computational limitations impede the use of significantly larger trap sizes. 

Considering the population dynamics in Fig.~\ref{fig:WM_popuQ}(a), we note that, as for resonant excitation, the population of the Rydberg level increases due to the dephasing, which also damps out the initial oscillations. The enhanced Rydberg population in the dissipative case is due to two distinct excitation mechanisms for the coherent and incoherent case as detailed above.
An indication for this different dynamics arising in off-resonant excitation is observed in the $Q$ parameter shown in Fig.~\ref{fig:WM_popuQ}(b). While in the resonant case, $Q\leq 0$ for all times, off-resonance the $Q$ parameter exhibits a maximum $Q>0$ at short times before decreasing to sub-Poissonian values (cf. inset). For larger systems, higher densities, and larger detunings, the super-Poissonian ($Q>0$) feature is much more pronounced \cite{schempp2013}, as discussed below.

\begin{figure}[t]
  \centering
 \includegraphics[width=\columnwidth]{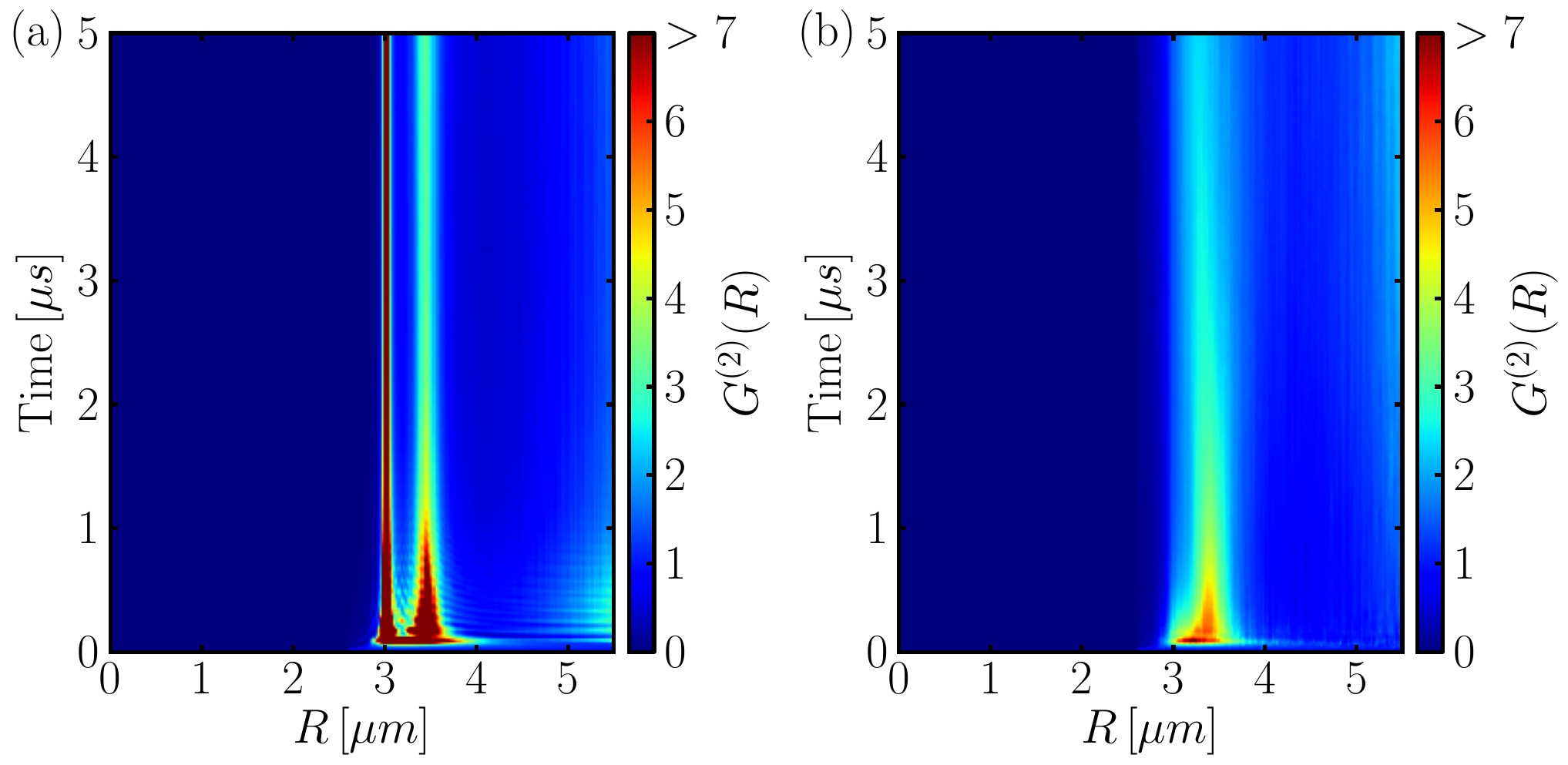}
 \caption{(Color online) Pair correlation function as a function of time for different dephasings, $\mathrm{(a)}$ $\Gamma/2\pi=0.05$\,MHz and $\mathrm{(b)}$ $\Gamma/2\pi=4$\,MHz. Further parameters as in Fig.~\ref{fig:WM_popuQ}. In (a), Rabi oscillations can be seen in the pair correlation function at small times.}
 \label{fig:WM_G2}
\end{figure}

To get a better understanding of the relevant processes contributing to off-resonant excitation dynamics, we consider again the pair correlation function, shown in Fig.~\ref{fig:WM_G2}. For the laser parameters used in the simulations, the two-photon resonance condition yields as resonance distance $R_{2\gamma}=[C_6/(2\Delta)]^{1/6}\approx 3.04\,\mu m$ while the single-photon resonance occurs at $R_\gamma=(C_6/\Delta)^{1/6}\approx 3.42\,\mu m$. For small dephasing, $\Gamma/2\pi=0.05$\,MHz, the two-photon resonance dominates, with the $G^{(2)}(R_{2\gamma})$ peak value being around one order of magnitude larger than the single-photon $G^{(2)}(R_{\gamma})$ value. For large dephasing, in contrast, the two-photon peak effectively disappears, and only a smeared-out single-photon peak is retained. This is a direct consequence of the dephasing, which can be intuitively understood by considering the damping effect of the dephasing on the inter-atomic coherences. In a two-atom picture, the direct transition $\ket{gg}\leftrightarrow\ket{rr}$ not populating the singly excited states is possible only by means of two-atom coherences, which get destroyed by a large dephasing. Consequently, a large dephasing leads to population dynamics that always involve the singly excited states, which, in turn, result in a relative enhancement of single-photon processes. Accordingly, there are two regimes with respect to dephasing: For small dephasing, two-photon resonance processes dominate, whereas for large dephasing single-photon resonances determine the excitation dynamics \cite{schempp2013}. Thus, the presence of characteristic features in the pair correlation function allows us to unambiguously distinguish between coherent and incoherent dynamics. 

\begin{figure}[t]
  \centering
 \includegraphics[width=\columnwidth]{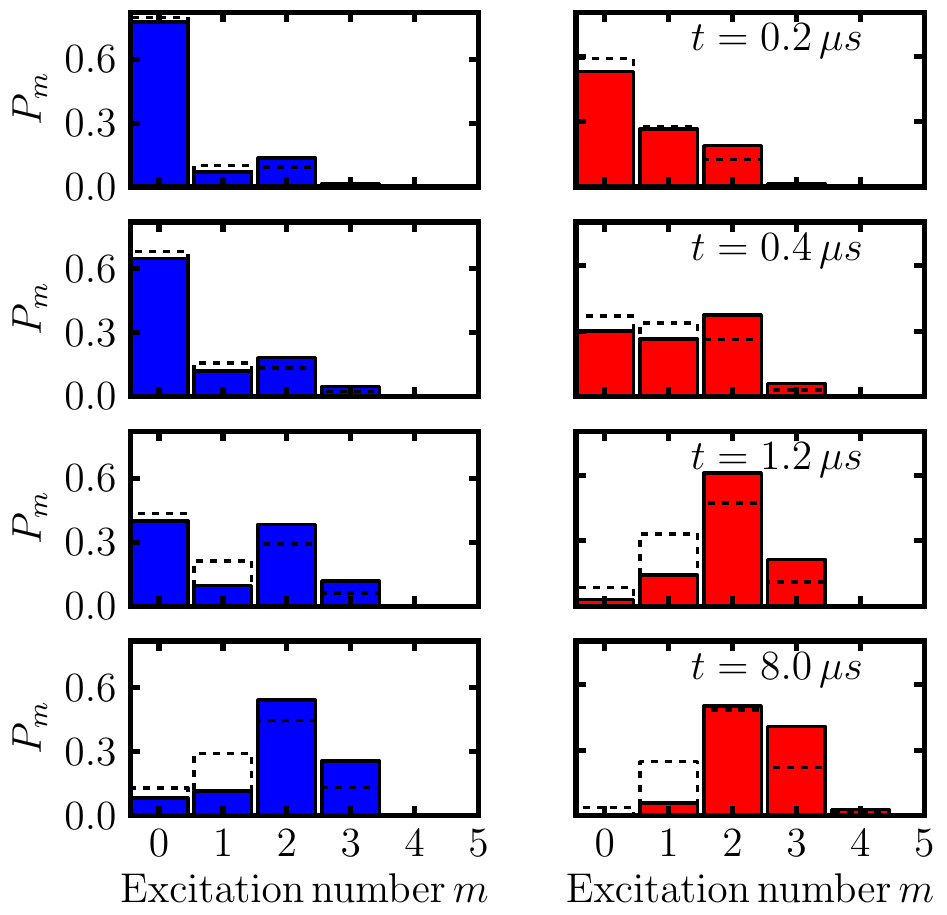}
 \caption{(Color online) Excitation probability $P_m$ for two different dephasings, $\Gamma/2\pi=0.05$\,MHz (blue, left panel) and $\Gamma/2\pi=4$\,MHz (red, right panel), at four different times. The dashed histograms denote the respective excitation statistics measured with a finite detection efficiency of $\eta=0.8$. Further parameters as in Fig.~\ref{fig:WM_popuQ}.}
 \label{fig:WM_histograms}
\end{figure}

The difference in the dynamics also manifests itself in the excitation number histograms in Fig.~\ref{fig:WM_histograms}. In the quasi-coherent case, a bimodality ($P_1<P_0,P_2$) arises at small times, since direct two-photon excitation ($m=0\leftrightarrow m=2$) governs the dynamics, leading to a strongly reduced probability to find a single excitation in the system. This can be directly read off the left panel of Fig.~\ref{fig:WM_histograms}, which shows that the singly excited state manifold $P_1$ is significantly suppressed with respect to the zero-fold ($P_0$) and doubly ($P_2$) excited state manifold at small times. The ground state population therefore migrates to the higher excited state manifolds via the $m=0\leftrightarrow m=2$ bimodality such that the singly excited state is only sparsely populated. When measuring the excitation histograms with a finite detection efficiency $\eta$, the bimodal structure gets smeared out (cf. \cite{schempp2013}). Still, a detection efficiency of $\eta=0.8$ is sufficient to resolve the bimodal structure (dashed lines in Fig.~\ref{fig:WM_histograms}) in our simulations.

\begin{figure}[t]
  \centering
 \includegraphics[width=\columnwidth]{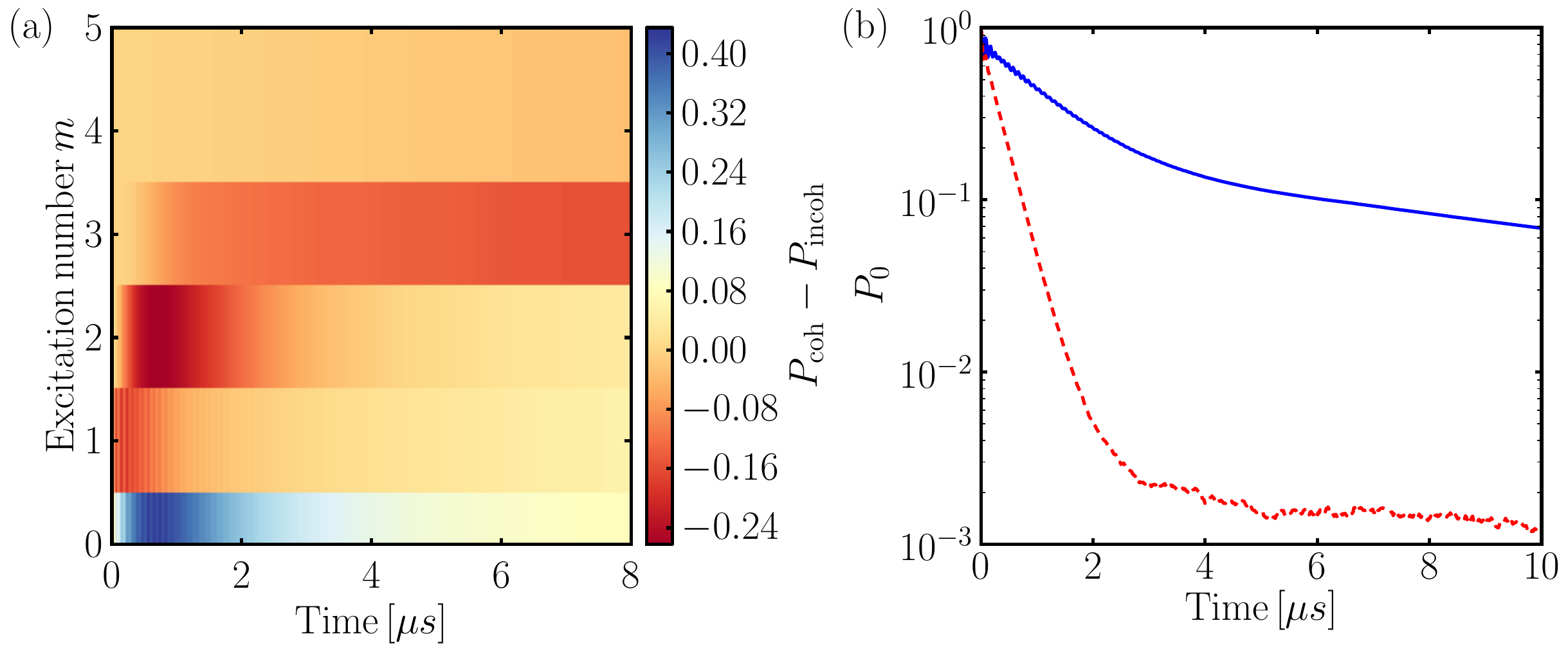}
 \caption{(Color online) $\mathrm{(a)}$ Difference between quasi-coherent ($\Gamma/2\pi=0.05$\,MHz) and incoherent ($\Gamma/2\pi=4$\,MHz) excitation histograms and $\mathrm{(b)}$ excitation probability $P_0$ on a logarithmic scale as a function of time. In $\mathrm{(b)}$, the solid blue curve corresponds to $\Gamma/2\pi=0.05$\,MHz whereas the dashed red one corresponds to $\Gamma/2\pi=4$\,MHz. Further parameters as in Fig.~\ref{fig:WM_popuQ}.}
 \label{fig:WM_P0uhistdiff}
\end{figure}

In the strongly dissipative case ($\Gamma/\Omega = 4$), this kind of bimodality does not arise. Rather, we find sequential growth of excitations around the first off-resonant excitation (acting as an initial seed) \cite{schempp2013}. The sequential growth is due to single-photon resonance processes and consequently populates the $m=1$ subspace appreciably (cf. right panel of Fig.~\ref{fig:WM_histograms}).
This can also be seen from Fig.~\ref{fig:WM_P0uhistdiff}(a), where the difference between the two excitation histograms, $P_m(\Gamma/2\pi=0.05\,\mathrm{MHz})-P_m(\Gamma/2\pi=4\,\mathrm{MHz})$, is shown. Already after a small evolution time, a population excess in the $P_{>0}$ subspaces compared to the quasi-coherent case emerges in the presence of large dephasing [red (dark grey) color in Fig.~\ref{fig:WM_P0uhistdiff}(a)]. The dynamics of the incoherent population excess indicate that in incoherent excitation the population moves via the single excitation subspace rapidly towards larger excitation numbers as compared to the quasi-coherent case, where a comparably large fraction of the population is trapped in the ground state for several $\mu s$.
This is illustrated in Fig.~\ref{fig:WM_P0uhistdiff}(b), which shows the ground state population $P_0$ on a logarithmic scale for both dephasing constants. Evidently, the ground state depletion significantly speeds up in the presence of strong dephasing. The timescale for the ground state depletion depends on the detuning value \cite{schempp2013} as well as on the density and the trap geometry. The trade-off between the timescales of the ground state depletion can be understood by noting that in the coherent case the characteristic timescale is the two-photon Rabi coupling $\Omega^2/\Delta$, enhanced by the number of available resonant atom pairs which gets small at large detuning, while in the incoherent case the timescale is determined by the off-resonant excitation rate of single atoms enhanced by the number of all atoms $N$ in the sample.  

\subsection{\label{subsec:largetrap}Large trap geometry}
In a larger trap, the single-photon dominated excitation process emerging in the presence of large dephasing can induce interesting features in the excitation dynamics. Due to the computational limitations of the MCWF technique with respect to the number of many-body states and consequently the number of excitations, we need to switch to another model, namely the rate equation \cite{ates2007}, for the discussion of larger trap geometries.

\subsubsection{\label{subsec:RE_benchmark}Justification of the rate equation treatment}
The basic idea behind the rate equation model already introduced in Sec.~\ref{subsec:offres_mechanisms} is to reduce the many-body master equation to a rate equation for the populations of the many-body states by adiabatically eliminating the single-atom coherences, and to take into account the Rydberg-Rydberg interaction via an effective energy shift of the Rydberg level only. 
As a consequence of the classical treatment, Rydberg systems consisting of several thousands of atoms can be simulated.
In this section, we briefly discuss a two-level rate equation benchmark, comparing MCWF calculations with rate equation results, and show that rate equation treatment is valid in the presence of strong decoherence (dephasing).
The rate equation we use is essentially the one introduced in Ref.~\cite{ates2007}.

\begin{figure}[t]
  \centering
 \includegraphics[width=\columnwidth]{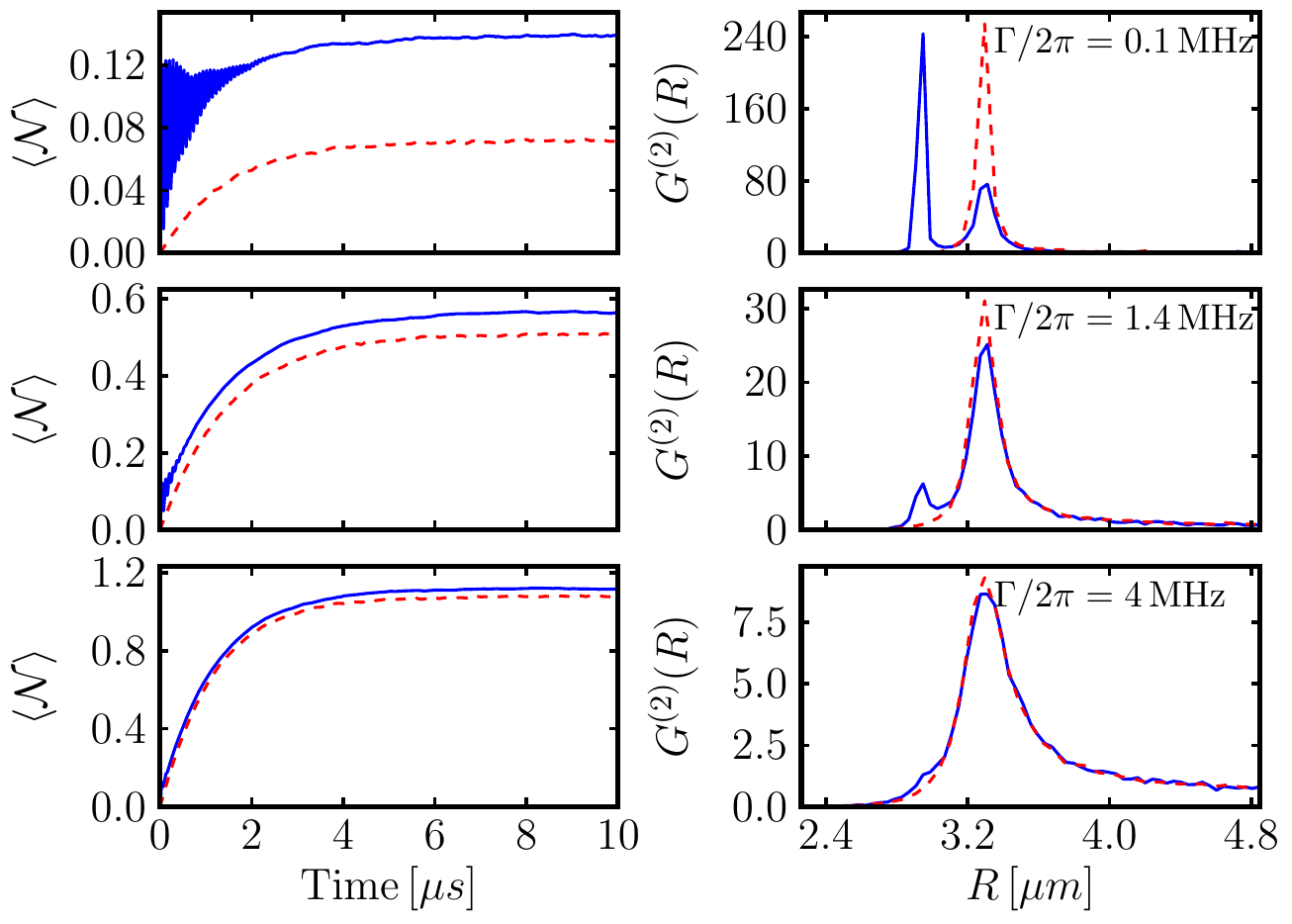}
 \caption{(Color online) $\mathrm{(a)}$ Mean Rydberg population $\langle\mathcal{N}\rangle$ as a function of time and $\mathrm{(b)}$ pair correlation function, comparing MCWF model (blue, solid) and rate equation model (red, dashed) for different dephasings. The pair correlation function is evaluated at $t=10\,\mu s$.  To improve visibility, only the lowest order peaks of the pair correlation function are shown. Further parameters: $N=30$, $C_6/2\pi=16\,\mathrm{GHz}\,\mu m^6$, $\Omega/2\pi=0.8$\,MHz, $\gamma/2\pi=0.125$\,MHz, $\Delta/2\pi = 12$\,MHz, one-dimensional trap of length $L_\mathrm{1D}=16.2\,\mu m$.}
 \label{fig:kMC_benchmark}
\end{figure}

Considering Fig.~\ref{fig:kMC_benchmark}, which shows both mean Rydberg population $\langle\mathcal{N}\rangle$ and pair correlation function $G^{(2)}(R)$ for different dephasings, we find that rate equation and MCWF calculations agree for large dephasing, but deviate strongly in the coherent regime (cf. also \cite{petrosyan2013a,schempp2013}). This is to be expected, since in the rate equation derivation, inter-atomic coherences and thus multi-photon processes are ignored as the Rydberg interaction is incorporated  via an energy shift in the detuning of the individual atoms \cite{ates2007} only. In fact, the pair correlation function illustrates how the approximations introduced during the derivation of the rate equation are validated in the presence of strong dephasing, while being totally inappropriate in quasi-coherent excitation.
That is, two-photon resonance processes quantified by the peak at $[C_6/(2\Delta)]^{1/6}\approx 2.96\,\mu m$ are not included in the rate equation model, which captures only single-photon resonance effects [peak at $(C_6/\Delta)^{1/6}\approx 3.32\,\mu m$]. 
Moreover, the dynamics of the system is only followed approximately, i.e., the rate equation does not reproduce the (damped) oscillations in the dynamics of the population or the $Q$ parameter. 

The systematically underestimated Rydberg population in the rate equation model compared to MCWF calculations implies that two-photon effects contribute to the population (e.g. via seeding excitations) even when two-photon resonance processes are suppressed with respect to single-photon processes. The relative error, however, decreases as the dephasing increases, from $\approx 49\%$ for $\Gamma/2\pi=0.1$~MHz to $\approx 3\%$ for $\Gamma/2\pi=4$~MHz.

\subsubsection{\label{subsec:large_trap}Results}
\begin{figure}[t]
  \centering
 \includegraphics[width=\columnwidth]{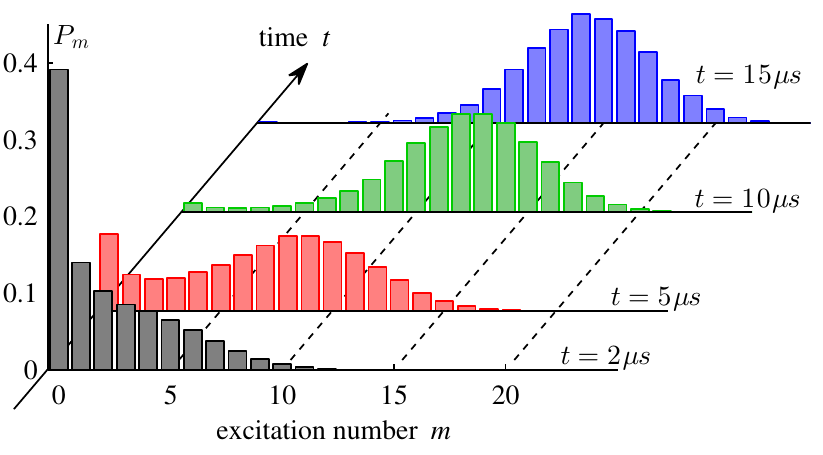}
 \caption{(Color online) Excitation probability $P_m$ at different times for the experimental trap geometry of Ref.~\cite{schempp2013}, $\Delta/2\pi = 15$~MHz and a density of $1.5\times 10^{12}$\,cm$^{-3}$, obtained using rate equation calculations \cite{schempp2013,ates2007}.}
 \label{fig:WM_RE_histtdep}
\end{figure}
Having shown that the rate equation is valid in the presence of large dephasing, we now consider a larger trap geometry, i.e., the setup of Ref.~\cite{schempp2013}, which is also off-resonantly driven and strongly dissipative, i.e., single-photon processes dominate.
In a larger trap, a bimodal structure can arise in the excitation dynamics, but for completely different reasons than in the case of small dissipation (cf. Sec.~\ref{subsec:smalltrap}). For strong dissipation, the bimodal feature that is observed is due to the two timescales involved in the excitation process, namely the timescale of the ground state depletion [cf. Fig.~\ref{fig:WM_P0uhistdiff}(b)], which slows down with increasing detuning \cite{schempp2013}, and the timescale of the resonant single-photon excitation $\gamma_{\uparrow,\mathrm{res}}$.
If the timescale of the ground state depletion is significantly smaller than the timescale of the resonant excitation, a substantial fraction of the population is trapped in the ground state  $\ket{gg\cdots g}$ while the remaining population is spread over the manifold of states with multiple excitations.

Employing the experimental setup of Ref.~\cite{schempp2013} as well as the density $1.5\times 10^{12}$\,cm$^{-3}$ and $\Delta/2\pi = 15$~MHz, we find for the dynamics of the excitation histograms the one shown in Fig.~\ref{fig:WM_RE_histtdep}. At $t=5\,\mu s$, a bimodal distribution is clearly visible, which is due to the fast single-photon resonance processes which occur on a timescale much faster than the ground state depletion, leading to a rapid population of the higher excitation number subspaces. The broad and bimodal excitation histograms entail a super-Poissonain $Q$ value \cite{schempp2013}.
We emphasize, however, that this is a transient feature. At large times, the bimodal structure vanishes as the histograms peak around the mean count, leading to sub-Poissonian ($Q<0$) statistics again.

\subsubsection{\label{subsec:bimodalities}Discussion of the two bimodalities}
The reason that this kind of bimodality and the super-Poissionian $Q$ parameter values entailed by it are hardly visible in Fig.~\ref{fig:WM_histograms} and Fig.~\ref{fig:WM_popuQ}(b), respectively, is that our numerical calculations using MCWF technique can only be applied to small system sizes, in which the $m=2$ subspace is naturally populated. Thus, the two bimodalities cannot be unequivocally told apart, due to the clear limitations of the maximal excitation number. 
In addition, the ground state depletion, whose timescale increases for increasing detuning \cite{schempp2013}, is for $\Delta/2\pi = 10$\,MHz not yet slow enough to allow for significant long-term dynamics, so the interesting features in the excitation dynamics occur only during a small time interval in the setup considered for MCWF calculations.

Note that the long-term shape of Fig.~\ref{fig:WM_histograms} might not be unique to incoherent excitation. In the coherent case, however, the dynamics in the few-excitation manifold would be different regarding the dominance of two-photon resonance effects. Other than in the strongly dissipative case, an inversion between the singly and double excited states ($P_1<P_2$) would occur. Moreover, the overall timescale of the dynamics would differ, as seen in Fig.~\ref{fig:WM_popuQ}, since coherent higher-order processes occur on a larger timescale as compared to resonant single-photon resonance processes \cite{schempp2013}.

The visibility of the effects discussed above depends on both detuning, density and trap geometry. Since the width of the two-photon resonance decreases with increasing detuning \cite{gaerttner2013a}, a high density is required to observe strong two-photon effects in far off-resonant excitation. 
The single-photon resonance, conversely, being broader than the two-photon resonance in the first place, is further broadened by strong dephasing, even though damped with respect to the maximal value in quasi-coherent excitation. Thus, for strong resonance effects to be visible, many pairs need to feature distances that lie in the resonance width, which also requires large density and/or trap dimensionality. The time at which the interesting features in the excitation statistics can be observed depends further on the detuning chosen for off-resonant excitation.

\section{\label{sec:summary}Summary and discussion}
In summary we have shown that dephasing alters the excitation dynamics in Rydberg systems. While in resonant excitation, dephasing mainly increases the mean Rydberg population $\langle\mathcal{N}\rangle$ and lowers the $Q$ parameter value, its impact on off-resonant excitation dynamics was found to be more drastic. Since strong dephasing suppresses two-photon resonance processes, the dynamics in strongly dissipative systems was found to be rather determined by sequential single-photon processes than direct two-photon excitation of pairs of atoms. This could be directly quantified via the pair correlation function $G^{(2)}(R)$, which features two pronounced peaks for quasi-coherent laser driving, with the peak associated with two-photon resonance processes dominating by around one order of magnitude. For strong dephasing, only the peak associated with single-photon resonance processes is retained.

Studying the excitation number histograms in detail, we found qualitatively different dynamics arising in off-resonant excitation, each pertaining to the distinct regimes of dissipation. As a consequence of the two-photon resonance, a bimodality involving the ground and the twofold excited state was found for weak dephasing, with a strongly reduced population probability of the singly excited states. For strong dephasing, single-photon resonance effects that are fast compared to the timescale of the ground state depletion dominate the dynamics, allowing for another bimodal feature in the excitation histograms, characterized by appreciable ground state population and at the same time appreciable population of large excitation numbers. Both of the features are of transient nature and are accessible via time-resolved measurement of the excitation statistics.

Experimentally, the distinct dynamics arising in off-resonant excitation in the coherent and incoherent regime, respectively, could be discriminated by measuring either the pair correlation function $G^{(2)}(R)$ or the short-time dynamics of the excitation histograms.
Time-resolved measurement of the singly excited state manifold $P_1$ could provide insight into the dominant excitation mechanism present in the system, since the strong bimodality between zero and two excitations ($P_1<P_0,P_2$)  caused by the dominance of two-photon resonance processes is inherent in quasi-coherent excitation only.

In lattice geometries \cite{schauss2012}, where the two resonance processes can be toggled by choosing the appropriate detuning for a given lattice constant, the two kinds of excitation dynamics could be studied via time-dependent measurement of the excitation statistics, which could shed light on the differences in excitation dynamics between lattice and disordered \cite{schempp2013} geometries. This would be of particular intrest considering the $P_0, P_2$ bimodality at short times.

Hence, off-resonant excitation promises easier means to discern coherent and dissipative dynamics due to the different effect of dephasing on the single-photon and two-photon resonance, respectively.
\newline

\begin{acknowledgments}
We gratefully acknowledge helpful discussions with D. Breyel, K.\ Heeg, H.\ Schempp, G.\ G\"unter, C.\ S.\ Hofmann, A. Komnik, M.\ Robert-de-Saint-Vincent, M.\ Weidem\"uller, and S.\ Whitlock.
\end{acknowledgments}

%

\end{document}